\documentclass[runningheads]{llncs}

\usepackage{amsmath,graphicx}

\usepackage{soul}
\usepackage{bm}
\usepackage[textsize=small]{todonotes}

\usepackage{hyperref}
\usepackage{lipsum}
\usepackage{array,multirow,booktabs}
\usepackage{nccmath}
\usepackage{capt-of}

\def\x{{\mathbf x}}
\def\xlr{{\mathbf{x}^{\mathrm{LR}}}}
\def\xlri{{\mathbf{x}^{\mathrm{LR,i}}}}

\def\xhr{{\mathbf{x}^{\mathrm{HR}}}}
\def\xhatr{{\mathbf{\hat{x}}^{\mathrm{HR}}}}

\makeatletter
\def\blfootnote{\gdef\@thefnmark{}\@footnotetext}
\makeatother

\begin{document}
\title{Simulation-based parameter optimization for fetal brain MRI super-resolution reconstruction\thanks{This work is supported by the Swiss National Science Foundation through grants 182602 and 141283, and by the Eranet Neuron MULTIFACT project (SNSF 31NE30\_203977). We acknowledge access to the facilities and expertise of the CIBM Center for Biomedical Imaging, a Swiss research center of excellence founded and supported by CHUV, UNIL, EPFL, UNIGE and HUG.}}
\titlerunning{Simulation-based parameter optimization for fetal brain MRI SRR}

\author{Priscille de Dumast\inst{\dagger,1,2} \and Thomas Sanchez\inst{\dagger,1,2} \and Hélène Lajous\inst{1,2} \and Meritxell Bach Cuadra\inst{1,2}}

\authorrunning{P. de Dumast, T. Sanchez et al.}

\institute{Department of Radiology, Lausanne University Hospital (CHUV) and University of Lausanne (UNIL), Lausanne, Switzerland \and CIBM Center for Biomedical Imaging, Switzerland}
\blfootnote{\hspace{-.25cm}$^\dagger$ Priscille de Dumast and Thomas Sanchez contributed equally to this work.}

\maketitle
\begin{abstract}
  Tuning the regularization hyperparameter $\alpha$ in inverse problems has been a longstanding problem. This is particularly true in the case of fetal brain magnetic resonance imaging, where an isotropic high-resolution volume is reconstructed from motion-corrupted low-resolution series of two-dimensional thick slices. Indeed, the lack of ground truth images makes challenging the adaptation of $\alpha$ to a given setting of interest in a quantitative manner.
  In this work, we propose a simulation-based approach to tune $\alpha$ for a given acquisition setting. We focus on the influence of the magnetic field strength and availability of input low-resolution images on the ill-posedness of the problem. Our results show that the optimal $\alpha$, chosen as the one maximizing the similarity with the simulated reference image, significantly improves the super-resolution reconstruction accuracy compared to the generally adopted default regularization values, independently of the selected pipeline. Qualitative validation on clinical data confirms the importance of tuning this parameter to the targeted clinical image setting.
\end{abstract}
\section{Introduction}
\label{sec:introduction}

Magnetic resonance imaging (MRI) has become an increasingly important tool to investigate prenatal equivocal neurological situations, as it provides excellent anatomical details~\cite{griffiths2017use,enso2020role}. However, three-dimensional (3D) high-resolution (HR) imaging of the fetal brain is unfeasible due to the unpredictable fetal motion. In clinical practice, T2-weighted (T2w) fast spin echo (FSE) sequences are commonly used to minimize the effects of intra-slice random fetal movements by acquiring low-resolution (LR) series of two-dimensional (2D) thick slices \cite{gholipour2014fetal}. Nevertheless, the strong anisotropy of the images leading to partial volume effects on small structures within the fetal brain and the remaining inter-slice motion hamper the accurate analysis of 3D imaging biomarkers.

Several post-processing techniques have been proposed to combine multiple motion-corrupted LR series and leverage the information redundancy from orthogonal orientations to reconstruct a single, 3D isotropic HR motion-free volume of the fetal brain \cite{kuklisova-murgasova_reconstruction_2012,tourbier_efficient_2015,ebner_automated_2020,uus2022retrospective}. These approaches all feature several pre-processing steps (e.g. brain extraction, intensity correction, and harmonization) leading to slice-to-volume registration (SVR) where inter-series inter-slice motion is estimated, followed by super-resolution reconstruction (SRR). This latter step can be framed as an inverse problem of the form
\begin{equation}
  \min_{\x} \frac{1}{2} \| \mathbf{H}\x - \x^{\mathrm{LR}}\|^2+ \alpha R(\x),
  \label{eq:reconTV}
\end{equation}
where $\x$ is the target HR image, $\x^{\mathrm{LR}}$ the LR series, $\mathbf{H}$ an operator describing the motion, blurring and downsampling model estimated from the data, and $R$ the regularization function (e.g., total-variation (TV) \cite{tourbier_efficient_2015}, first-order Tikhonov \cite{ebner_automated_2020}, etc.). $\alpha$ is a parameter that balances the strength of the regularization term compared to the data fidelity term.

Various applications in medical image computing are formulated as inverse problems, and the optimization of regularization parameters has been widely studied in this context \cite{katsaggelos1992methods,afkham2021learning}. Most strategies explicitly rely on reference data, which are not available in the context of fetal MRI, making the setting of the regularization parameter $\alpha$ in a principled and quantitative manner highly challenging. To circumvent the lack of HR data of the fetal brain, several works use HR MR images from newborns as ground truth data and downsample them to simulate the acquisition of LR series that are then reconstructed and compared to the HR image to set the default value of their regularization parameters~\cite{kuklisova-murgasova_reconstruction_2012,tourbier_efficient_2015}. Alternative approaches consider a leave-one-out approach where the left-out LR series serves as a reference for the quantitative evaluation of the SRR~\cite{kuklisova-murgasova_reconstruction_2012,tourbier_efficient_2015}, or use a volume reconstructed from all available LR series as a reference to which SRR with fewer LR series can be compared \cite{ebner_automated_2020}. However, all of these works rely on constructing surrogate ground truths to study the influence of the regularization parameter on the quality of the SRR but do not provide insights on how to adapt it when new input acquisition setting has to be reconstructed. Furthermore, despite well-known differences in image acquisitions protocols, fetal brain SRR MRI studies are still carried out using the default regularization values of the selected pipeline~\cite{tourbier_efficient_2015,ebner_automated_2020,payette_automatic_2021,uus2022retrospective}.

This work proposes the first approach to optimize the setting of the regularization parameter $\alpha$ based on numerical simulations of imaging sequences tailored to clinical ones. We take advantage of a recent Fetal Brain MR Acquisition Numerical phantom (FaBiAN) \cite{lajous2022fetal} that provides a controlled environment to simulate the MR acquisition process of FSE sequences, and thus generates realistic T2w LR MR images of the fetal brain as well as corresponding HR volumes that serve as a reference to optimize the parameter $\alpha$ in a data-driven manner, considering both \textit{acquisition setting-specific} and \textit{subject-specific} strategies.

Our contributions are twofold. First, using synthetic, yet realistic data, we study the sensitivity of the regularization to three common variables in inverse SRR problem in fetal MRI: (i) the number of LR series used as input, (ii) the magnetic field strength which impacts also the in-plane through-plane spatial resolution ratio, and (iii) the gestational age (GA). Secondly, we qualitatively illustrate the practical value of our framework, by translating our approach to clinical MR exams. We show that $\alpha^*$ estimated by our simulated framework echoes a substantial improvement of image quality in the clinical SRR. To generalize the validity of our findings, we perform our study using two state-of-the-art SRR pipelines, namely MIALSRTK \cite{tourbier_efficient_2015} and NiftyMIC~\cite{ebner_automated_2020}.

\section{Materials and methods}\label{sec:materials}
\subsection{Simulated acquisitions}\label{ssec:simulation}
We use FaBiAN~\cite{lajous2022fetal,lajous_helene_2021_5599311} to generate T2w MR images of the developing fetal brain derived from a normative spatiotemporal MRI atlas (STA) ~\cite{gholipour_normative_2017}.
Typical FSE acquisitions are simulated using the extended phase graph (EPG) formalism, at either 1.5T or 3T, according to the MR protocol routinely performed at our local hospital for fetal brain examination. All sequence parameters are kept fixed at a given magnetic field strength (at $1.5/3T$: TR, $1200$/$1100 ms$; TE, $90$/$101 ms$; voxel size, $1.1\times 1.1\times 3$/$0.5\times 0.5\times 3 mm^3$)). Random complex Gaussian noise (mean, $0$; standard deviation, $0.15$ at 1.5T, respectively $0.0025$ at 3T) is added to the k-space data to qualitatively match the noise characteristics of clinical acquisitions. Multiple orthogonal LR series $\xlr = \{\xlri\}_i$ are simulated with a shift of the field-of-view of $1.6 mm$ in the slice thickness direction for series in the same orientation. The amplitude of fetal motion and the number of simulated LR series are further detailed in the experimental settings (Sections~\ref{ss:exp1} and ~\ref{ss:exp2}).
A reference HR isotropic volume $\xhr$ of the fetal brain is also simulated for each subject, without bias field or motion, to serve as a reference for the quantitative evaluation of the corresponding SRR. A visual comparison between clinical data is available at Figure~\ref{fig:fabian_clinical} in the Supplementary material.

\subsection{Super-resolution reconstruction methods}
Two widely adopted reconstruction pipelines, MIALSRTK~\cite{tourbier_efficient_2015} and NiftyMIC~\cite{ebner_automated_2020}, are used to reconstruct 3D isotropic HR images of the fetal brain from orthogonal LR series.
For each pipeline, we perform a grid search approach of the regularization parameter space.

\noindent\textbf{Remark.} Contrary to NiftyMIC~\cite{ebner_automated_2020}, MIALSRTK~\cite{tourbier_efficient_2015} places its regularization parameter $\lambda$ on the data fidelity term. For the sake of consistency, we will only use the formulation of Equation~\ref{eq:reconTV}, with $\alpha = 1/\lambda$ in the case of MIALSRTK.

\noindent\textbf{Quality assessment.} Solving Problem~\ref{eq:reconTV}
yields a SR-reconstructed image $\xhatr$ which quality can be compared against the reference $\xhr$ using various metrics. We use two common metrics for SRR assessment~\cite{kuklisova-murgasova_reconstruction_2012,tourbier_efficient_2015,ebner_automated_2020}, namely the peak signal-to-noise ratio (PSNR) and the structural similarity index (SSIM) \cite{wang2004image}.
The best regularization parameter $\alpha$ is identified as the one maximizing a given performance metric.

\subsection{Experiment 1 -- Controlled in silico environment}\label{ss:exp1}
In this first experiment, we study the sensitivity of the parameter $\alpha$ to common variations in the acquisition pipeline.

\noindent\textbf{Dataset.} For every STA subject, 9 LR series (3 per anatomical orientation) are simulated at 1.5T and 3T with little amplitude of stochastic 3D rigid motion\footnote{5\% of corrupted slices, translation of [-1,1]mm in every direction, 3D rotation of [-2,2]\textdegree}.

\noindent\textbf{Experimental setting.} We define four configurations based on the number of LR series given as input to the SRR pipeline (three or six series) and the magnetic field strength (1.5 or 3T). Note that the inter-magnetic field difference is especially captured in the image resolution, with a through-plane/in-plane ratio of $3.3/1.1 = 3$ at 1.5T and $3.3/0.5=6.6$ at 3T. In each configuration, individual brains are repeatedly reconstructed ($n=3$) from a selection of different LR series among the nine series available per subject.

The grid of parameters searched for NiftyMIC consists of 10 values geometrically spaced between $10^{-3}$ and $2$, plus the default parameter $\alpha_{\mathrm{def}} = 0.01$. For MIALSRTK, we use $\alpha \in \{1/0.75, \allowbreak 1/1.0, \allowbreak1/1.5,  \allowbreak1/2.0,  \allowbreak1/2.5,  \allowbreak1/3.0, \allowbreak 1/3.5, \allowbreak 1/5.0 \}$ ($8$ values, with default parameter $\alpha_{\mathrm{def}}=1/0.75$). At the end of the experiment, the best parameter, for either of the pipelines, is referred to as $\alpha_1^*$.

\noindent\textbf{Statistical analysis.} The optimal regularization parameters evaluated for the different SRR configurations are compared using the Wilcoxon rank sum test. The difference between the metrics performance obtained with default or optimal parameters is tested with a paired Wilcoxon rank sum test.
The $p$-value for statistical significance is set to $0.05$.

\subsection{Experiment 2 -- Clinical environment}\label{ss:exp2}
Clinical MR fetal exams are prone to substantial inter-subject variation and heterogeneity. In particular, the number of LR series available for reconstruction, as well as the amplitude of fetal motion, greatly vary from one subject to the other \cite{khawam_fetal_2021}.
Therefore, this second experiment has two purposes. First, we translate our findings from the first experiment to clinical data using the best value $\alpha_1^*$. Secondly, we study an alternative approach to perform a tailored subject-wise regularization tuning by simulating synthetic data for each subject that mimic the clinical acquisitions available. We refer to the obtained value as $\alpha_2^*$.

\noindent\textbf{Dataset.} Twenty fetal brain MR exams conducted upon medical indication were retrospectively collected from our institution. All brains were finally considered normal. Fetuses were aged between 21 and 34 weeks of GA (mean $\pm$ standard deviation (sd): $29.7 \pm 3.6$) at scan time. For each subject, at least three orthogonal series were acquired at 1.5T (voxel size: $1.125\times1.125\times3mm^3$). After inspection, four to nine series (mean $\pm$ sd: $6.3 \pm 1.5$) were considered exploitable for SRR.

The local ethics committee approved the retrospective collection and analysis of MRI data and the prospective studies for the collection and analysis of the MRI data in presence of a signed form of either general or specific consent.

The same 20 subjects are simulated using exam-specific parameters to mimic as closely as possible the corresponding clinical acquisitions. In particular, we match the number and the orientation of the LR series, as well as the amplitude of fetal motion (from little to moderate), and the GA of each subject.

\noindent\textbf{Experimental setting.} We consider the same regularization parameter space as in Experiment~1~(Section~\ref{ss:exp1}), and evaluate both clinical and simulated data on this parameter grid.

\noindent\textbf{Statistical analysis.} We compare the similarity between the images reconstructed by MIALSRTK and NiftyMIC using both default and optimized parameters. In this experiment, no reference images are available. Statistical significance of the performance difference is tested using a paired Wilcoxon rank sum test ($p<0.05$ for statistical significance).

\section{Results}
\subsection{Experiment 1 -- Controlled in silico environment}\label{ss:res_exp1}
\begin{figure}[!t]
  \centering
  \vspace{0pt}\includegraphics[width=\linewidth]{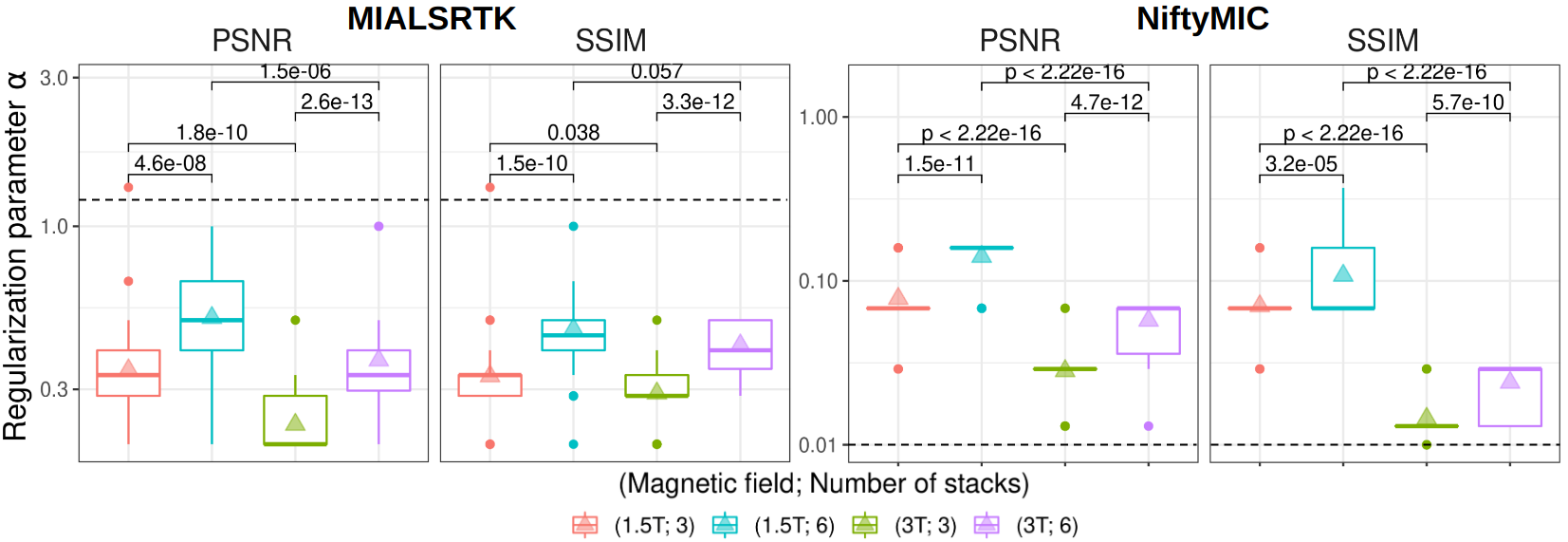}
  \captionof{figure}{Optimal regularization parameters $\alpha_1^*$ for MIALSRTK (left panel) and NiftyMIC (right panel) in the four different configurations studied. $\triangle$ indicates the mean optimal parameter. Inter-configurations $p$-values (Wilcoxon rank sum test, significance: $p<0.05$) are indicated. The dashed line shows the value of the default parameter for each SRR technique.}
  \label{fig:boxplot_magnetic_stack}
\end{figure}

\noindent\textbf{Optimal regularization parameter.}
Figure~\ref{fig:boxplot_magnetic_stack} shows the optimal regularization parameters $\alpha_1^*$ of SRR by MIALSRTK and NiftyMIC for each configuration.
Regardless of the magnetic field strength and the number of LR series used for reconstruction, the optimal regularization parameters that maximize the PSNR and SSIM compared to a synthetic HR volume greatly differ from the default values. For MIALSRTK, $\alpha_{\mathrm{def}}=1/0.75$, while the optimal range is found between $1/2.25$ and $1/4.5$. For NiftyMIC, $\alpha_{\mathrm{def}}=0.01$, whereas the optimal range is found between $0.015$ and $0.15$. 

We observe that for both the PSNR and the SSIM, the optimal regularization weight \textit{increases} with the number of series used in the reconstruction, and decreases with the resolution. This is because changing the number of LR series or the magnetic field strength affects the magnitude of the data fidelity term with respect to the regularization term. When more series are combined in the SRR, a larger regularization parameter must be used to keep the ratio $ \| \mathbf{H}\x - \x^{\mathrm{LR}}\|^2/\alpha R(\x)$ constant.

\noindent\textbf{Quality improvement.} The corresponding mean PSNR and SSIM values, computed across all subjects both with default and optimal regularization parameters, and compared to the reference HR volume are displayed in Table~\ref{tab:mean_performances}. Overall, the quality metrics obtained from the SRR with optimal parameters are significantly improved compared to those obtained with default values. These results strongly suggest that the regularization parameters are highly sub-optimally set for both SRR pipelines.
This is further illustrated on Figure~\ref{fig:simu_viz}, where we reconstruct simulated data and compare them to the corresponding ground truth image: using a default $\alpha$, MIALSRTK tends to overly smooth the image, while NiftyMIC reconstructs images that are artificially sharp, enhancing edges beyond what is present on the reference image.

\noindent\textbf{Gestational age-based analysis.} Since the human brain undergoes drastic morphological changes throughout gestation~\cite{tierney2009brain}, one could expect to adjust $\alpha$ to GA. However, our experiments (cf. Supp. Figure~\ref{fig:plot_ga}) suggest that the $\alpha^*$ does not depend on GA, and is in line with the values reported on Fig.~\ref{fig:boxplot_magnetic_stack}.

\begin{table}[!t]
  \centering
  \vspace{0pt}\captionof{table}{Mean metrics computed across all subjects on images reconstructed using the default regularization parameter $\alpha_{\mathrm{def}}$ or the optimal parameter $\alpha_1^*$ respectively, compared to the simulated reference HR volume, for the four configurations studied.
    $\dagger$ indicates paired Wilcoxon rank sum test statistical significance ($p<0.05$).}
  \label{tab:mean_performances}
  \resizebox{.6\linewidth}{!}{
    \begin{tabular}{ccccccccc}
      \toprule
                              & \multicolumn{4}{c}{MIALSRTK}          & \multicolumn{4}{c}{NiftyMIC}                                                                                                                                                                                                  \\
                              & \multicolumn{2}{c}{PSNR ($\uparrow$)} & \multicolumn{2}{c}{SSIM ($\uparrow$)} & \multicolumn{2}{c}{PSNR ($\uparrow$)} & \multicolumn{2}{c}{SSIM ($\uparrow$)}                                                                                                         \\
      \cmidrule(lr){2-3} \cmidrule(lr){4-5} \cmidrule(lr){6-7} \cmidrule(lr){8-9}
      (Field strength; \# LR) & $\alpha_{\mathrm{def}}$               & $\alpha_1^*$                          & $\alpha_{\mathrm{def}}$               & $\alpha_1^*$                          & $\alpha_{\mathrm{def}}$ & $\alpha_1^*$            & $\alpha_{\mathrm{def}}$ & $\alpha_1^*$            \\
      \midrule

      (1.5T; 3)               & 18.9                                  & \textbf{20.2}$^\dagger$               & 0.78                                  & \textbf{0.80}$^\dagger$               & 17.3                    & \textbf{20.8}$^\dagger$ & 0.79                    & \textbf{0.82}$^\dagger$ \\
      (1.5T; 6)               & 20.1                                  & \textbf{20.8}$^\dagger$               & 0.82                                  & \textbf{0.83}$^\dagger$               & 17.0                    & \textbf{21.5}$^\dagger$ & 0.80                    & \textbf{0.84}$^\dagger$ \\
      (3T; 3)                 & 19.9                                  & \textbf{21.8}$^\dagger$               & 0.75                                  & \textbf{0.77}$^\dagger$               & 20.5                    & \textbf{21.2}$^\dagger$ & 0.77                    & \textbf{0.77}$^\dagger$ \\
      (3T; 6)                 & 21.0                                  & \textbf{22.2}$^\dagger$               & 0.78                                  & \textbf{0.80}$^\dagger$               & 20.9                    & \textbf{22.0}$^\dagger$ & 0.80                    & \textbf{0.80}$^\dagger$ \\
      \bottomrule
    \end{tabular}}
\end{table}
\subsection{Experiment 2 -- Clinical environment}\label{ss:res_exp2}

In this experiment, we compare two differently optimized regularization parameters. First, we use the optimal value $\alpha_1^*$ (from Fig.~\ref{fig:boxplot_magnetic_stack} at 1.5T, and rounded to the closest value on the grid of parameters). Second, we use the optimized regularization parameter $\alpha_2^*$ estimated from the subject-specific simulation.

\begin{figure}[!t]
  \centering
  \vspace{0pt}\includegraphics[width=.85\linewidth]{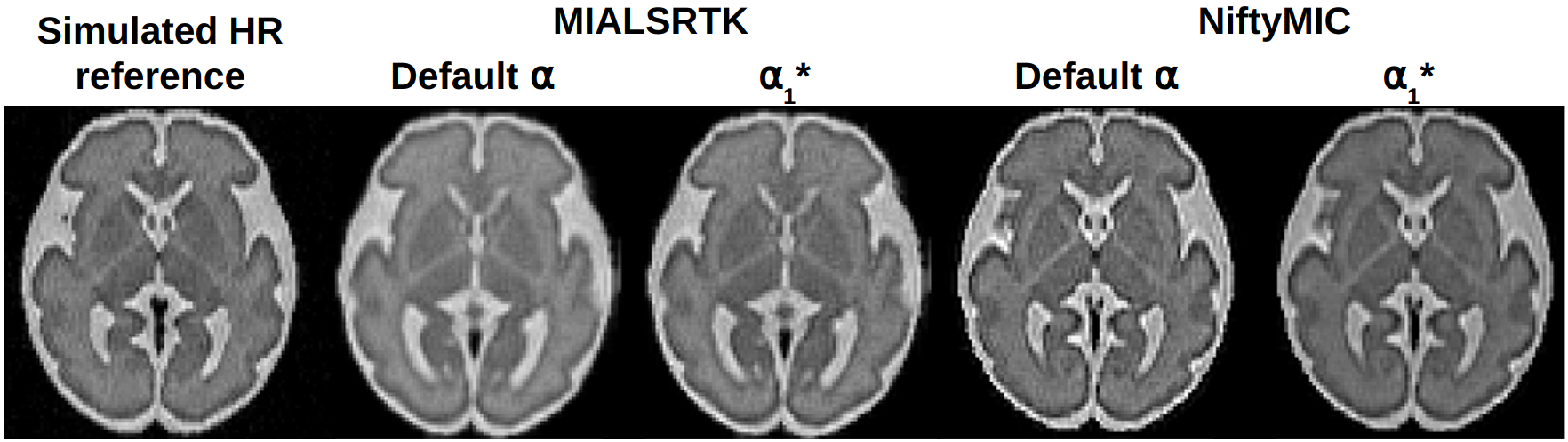}
  \caption{Axial comparison of SRR simulated cases of a 30-week old subject, using default and optimal parameters from Experiment~1 ($\alpha_1^*$), reconstructed with four LR series.}
  \label{fig:simu_viz}

  \centering
  \includegraphics[width=0.9\linewidth]{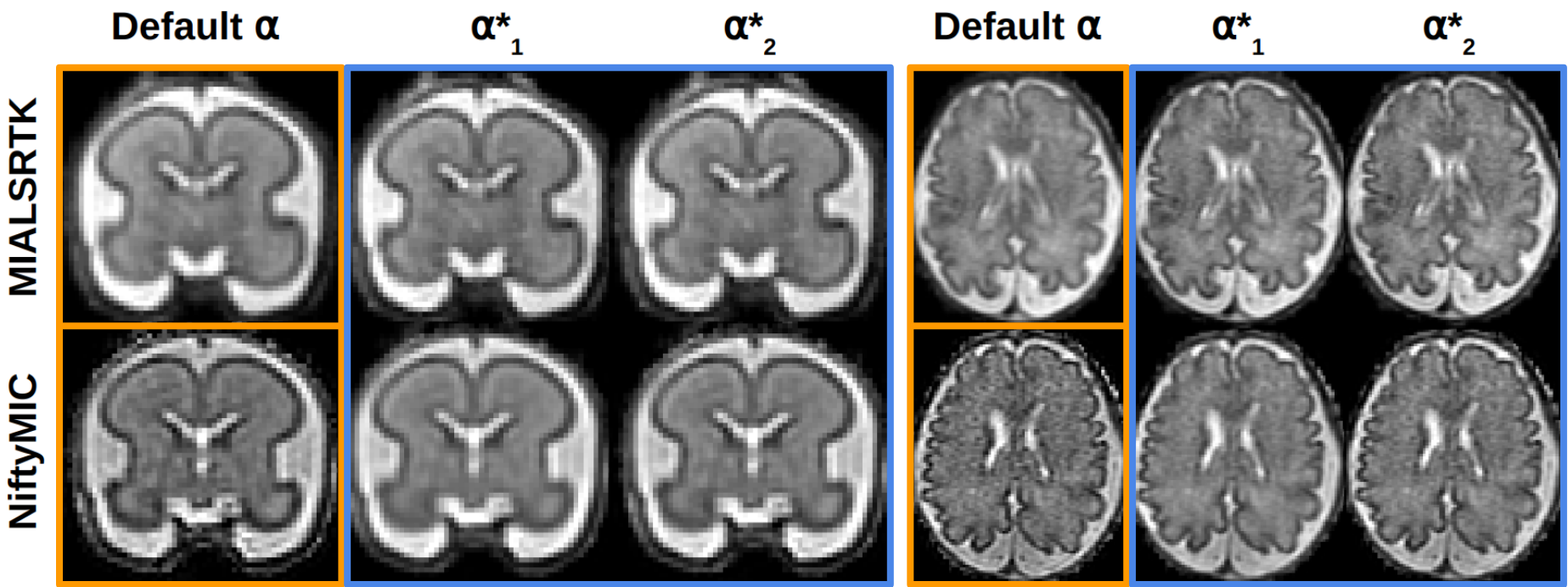}
  \caption{Illustration of domain shift, on two SRR of clinical subjects.The SR reconstructions using the default $\alpha$ (orange boxes) are much more different than the ones reconstructed with the optimal parameters (blue boxes).}
  \label{fig:clinical_vis}

  \includegraphics[width=0.65\linewidth]{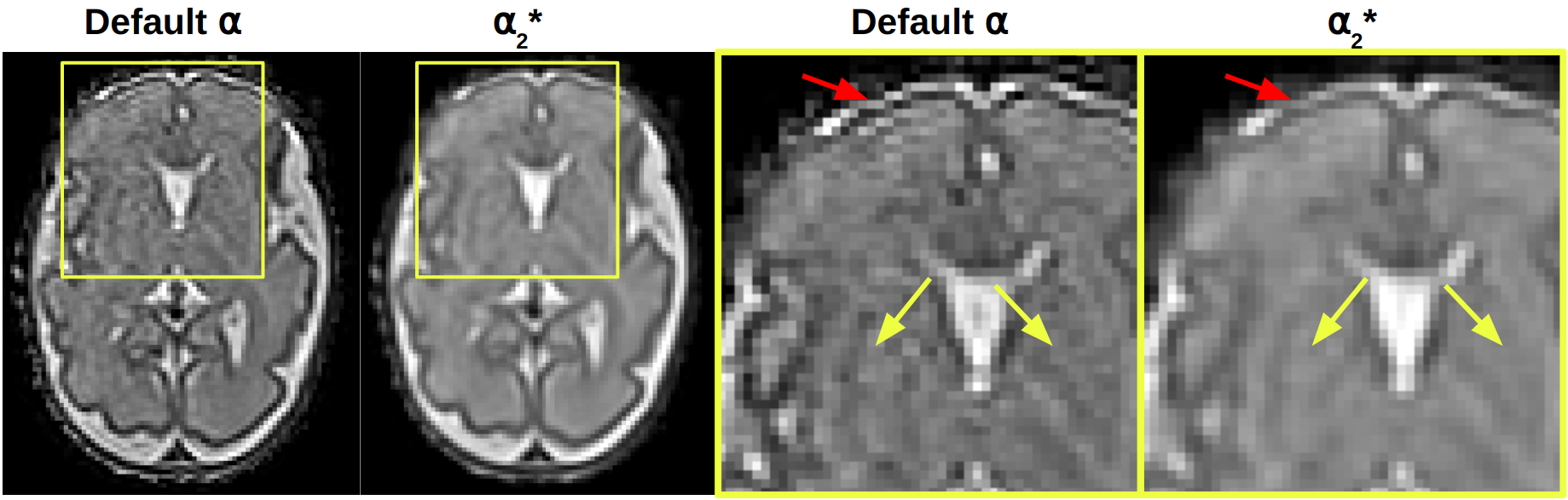}
  \caption{Comparison of two clinical cases reconstructed using NiftyMIC with different regularization parameters. We see that the optimal parameter $\alpha_2^*$ yields an image with less ringing artifacts at the frontal lobe (red arrow) and a more clearly delineated deep gray matter (yellow arrows).}\label{fig:dgm}
\end{figure}

Figure~\ref{fig:clinical_vis} shows the SRR of two subjects with default and optimal parameters using both pipelines. We observe that using the optimal parameters $\alpha^*_1$ and $\alpha^*_2$ makes the reconstructed images more similar.
Indeed, the default parameters of MIALSRTK and NiftyMIC promote opposite behaviors, towards smoother (i.e., the default regularization is higher than the optimal one), respectively noisier (the default regularization is lower than the optimal one) images.

\begin{figure}[!t]
  \centering
  \begin{minipage}[t]{0.44\textwidth}
    \vspace{0pt}\includegraphics[width=\linewidth]{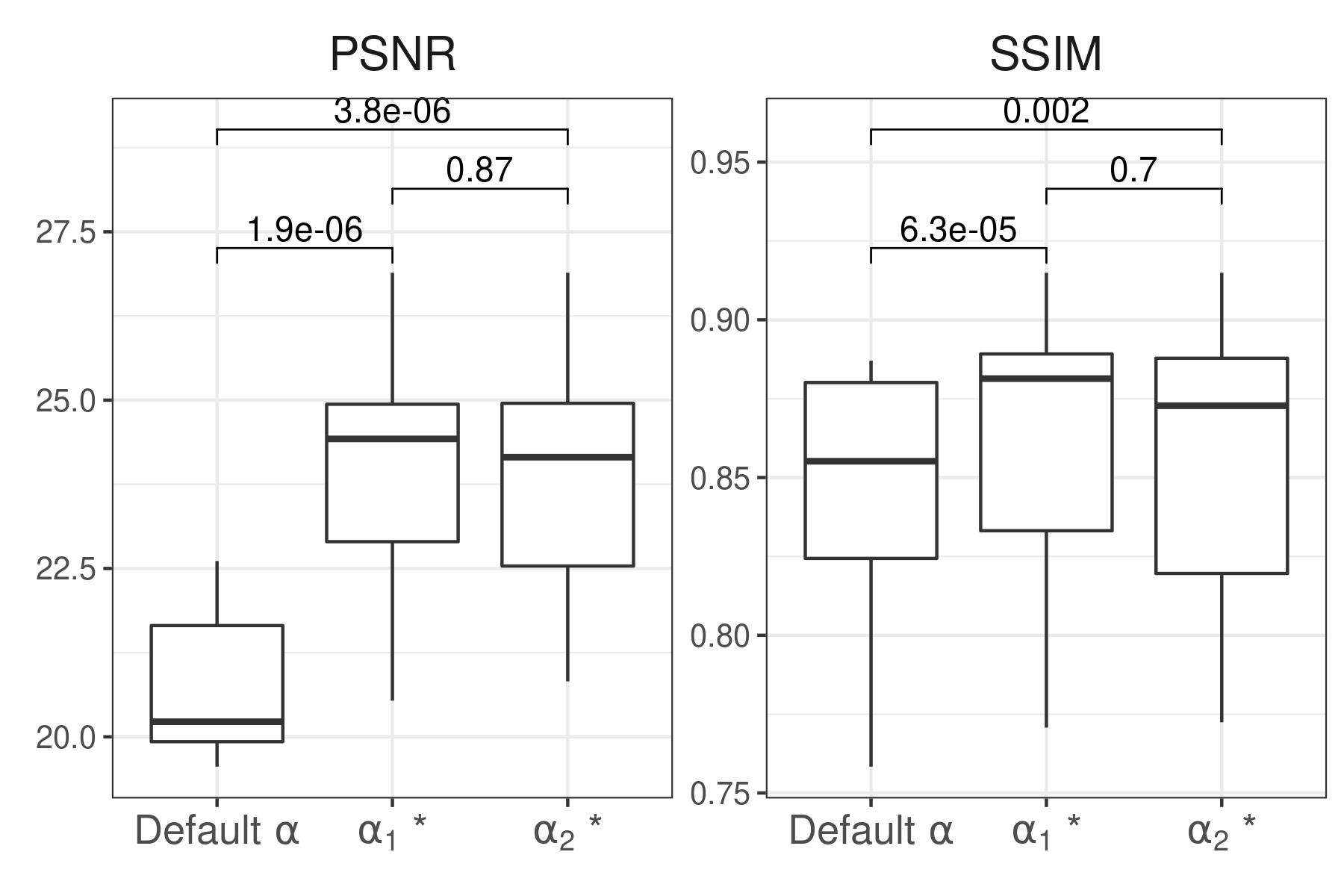}
    \caption{Similarity (PSNR and SSIM) between the images reconstructed using MIALSRTK and NiftyMIC with default and optimal regularization parameters. Comparison is done between all 20 clinical exams; $p$-values from paired Wilcoxon rank sum test, statistical significance: $p< 0.05$).}
    \label{fig:inter_srr_boxplot}
  \end{minipage}
  \hfill
  \begin{minipage}[t]{0.54\textwidth}
    \vspace{0pt}\includegraphics[width=\linewidth]{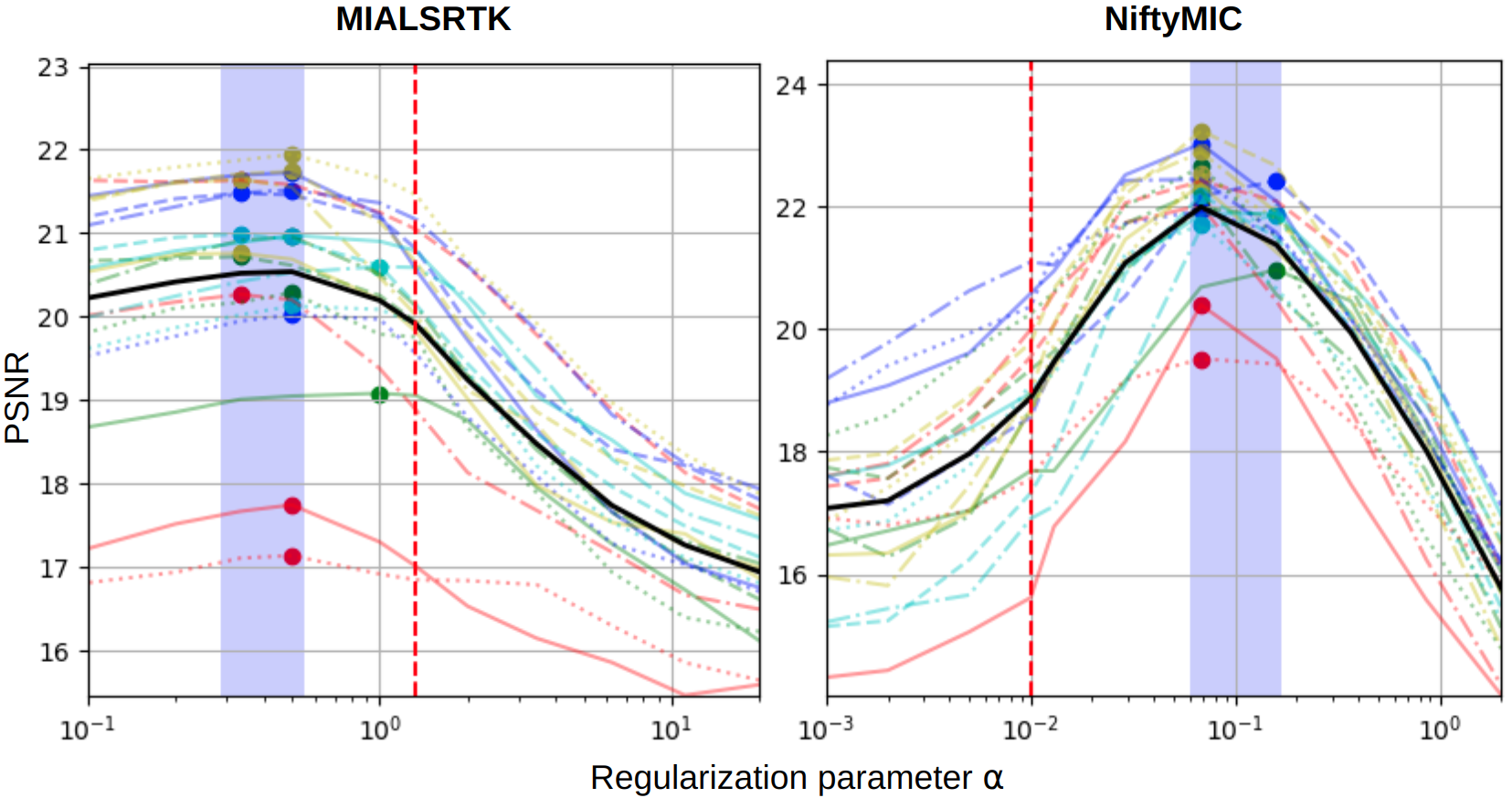}
    \caption{PSNR (left: MIALSRTK,  right: NiftyMIC) for the simulated subjects of Experiment~2. Every light line represents an individual subject, and their average is the bold black curve. Dashed red (vertical) line is the default regularization for each method. Purple region highlights the range of optimal parameters determined in Experiment~1 at 1.5T.}
    \label{fig:subject_wise_param}
  \end{minipage}
\end{figure}

We quantitatively confirm the similarity of the optimized SRR by computing the PSNR and the SSIM between the reconstructed images from both methods 
for $\alpha_{\textrm{def}}$, $\alpha^*_1$ and $\alpha^*_2$. The results are shown in Figure~\ref{fig:inter_srr_boxplot}. The difference between the default and optimized parameters is statistically significant for both metrics. There is however no significant difference between the images reconstructed using the parameters optimized \textit{setting-wise} ($\alpha_1^*$) and \textit{subject-wise} ($\alpha_2^*$). As shown in Fig.~\ref{fig:subject_wise_param}, the parameters $\alpha_2^*$ optimized based on subject-specific simulations always lie within the range of optimal parameters $\alpha_1^*$ determined in Exp.~1.

Beyond more similar images, optimizing the regularization parameter can also matter in terms of the structures that will be visible on the image. On Figure~\ref{fig:dgm}, using the optimal $\alpha_1^*$ allows to delineate the deep gray matter more clearly compared to the $\alpha_{\text{def}}$.

\section{Discussion and Conclusion}
In this paper, we propose a novel simulation-based approach that addresses the need for automated, quantitative optimization of the regularization hyperparameter in ill-conditioned inverse problems, with a case study in the context of SRR fetal brain MRI. Our estimated regularization weight shows both qualitative and quantitative improvements over widely adopted default parameters. Our results also suggest that \textit{subject-specific} parameter tuning -- which is computationally expensive to run -- might not be necessary, but that an \textit{acquisition setting-specific} tuning, ran only once, might be sufficient in practice.

As such, the proposed methodology demonstrates a high practical value in a clinical setting where fetal MR protocols are not standardized, leading to heterogeneous acquisition schemes across centers and scanners.
Besides, we show that our simulation-based optimization approach reduces the variability in image quality and appearance between the two SRR pipelines studied. We expect this behavior to contribute to mitigating the domain shift currently inherent to any reconstruction technique, a key challenge in the development of automated tissue segmentation methods~\cite{payette_efficient_2020,dedumast_syntheticMRI_2022}.
Future work will address some limitations of this study. Indeed, we mostly focused on the influence of the main magnetic field strength and the number of available LR series, but the signal-to-noise ratio within LR series may also affect the regularization setting~\cite{lajous2022fetal}. This aspect could also be tuned within the proposed MR acquisition simulation framework. Moreover, clinical assessment by radiologists of the different SRR would be important to further validate our method. Such an evaluation would allow to compare our approach to other techniques for parameter tuning, which cannot be done quantitatively due to the lack of HR ground truth data.

\bibliographystyle{IEEEtran}
\bibliography{refs}

\appendix
\newpage
\section*{Supplementary material}

\begin{figure}[!h]
  \centering
  \includegraphics[width=\linewidth]{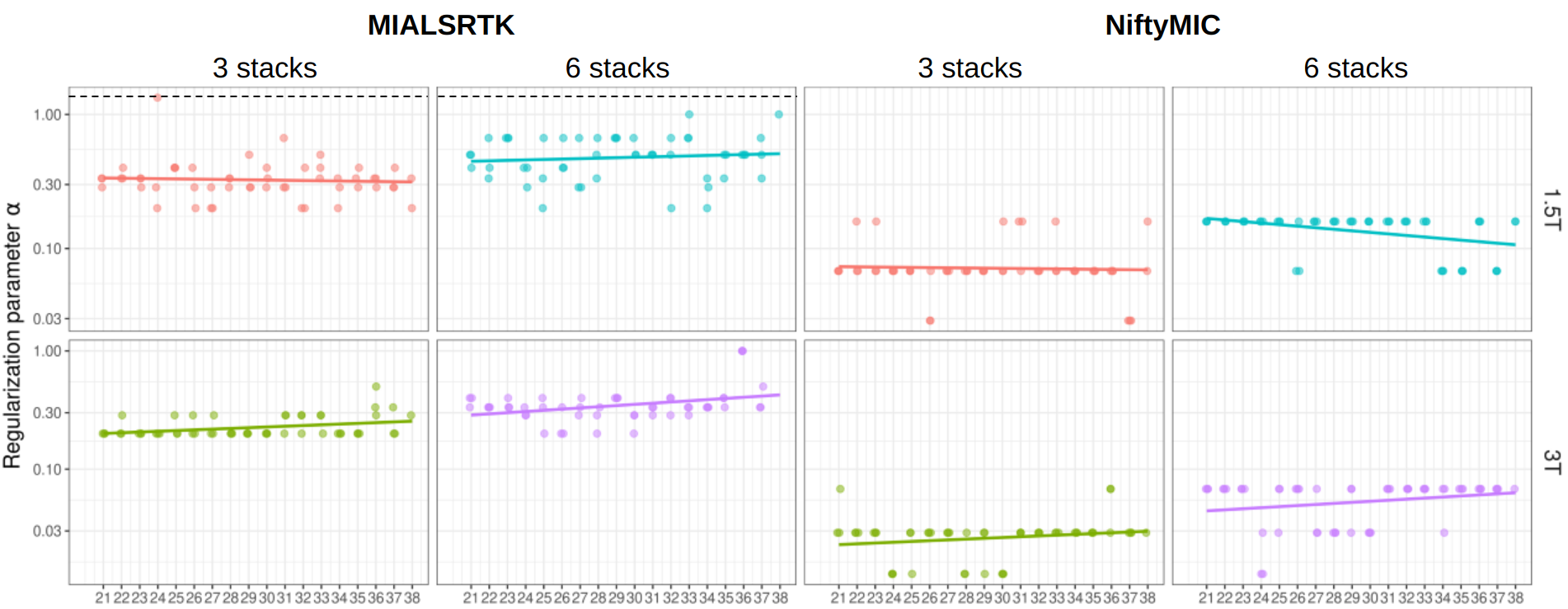}
  \caption{Optimal parameters for MIALSRTK and NiftyMIC as a function of the gestational age (GA). There is only a very moderate effect at best on the optimal regularization parameter as a function of the GA. The default parameter for MIALSRTK is $\alpha=4/3$ (dashed line) and for NiftMIC, $\alpha = 0.01$ is below the y-axis.}
  \label{fig:plot_ga}
  \includegraphics[width=.97\linewidth]{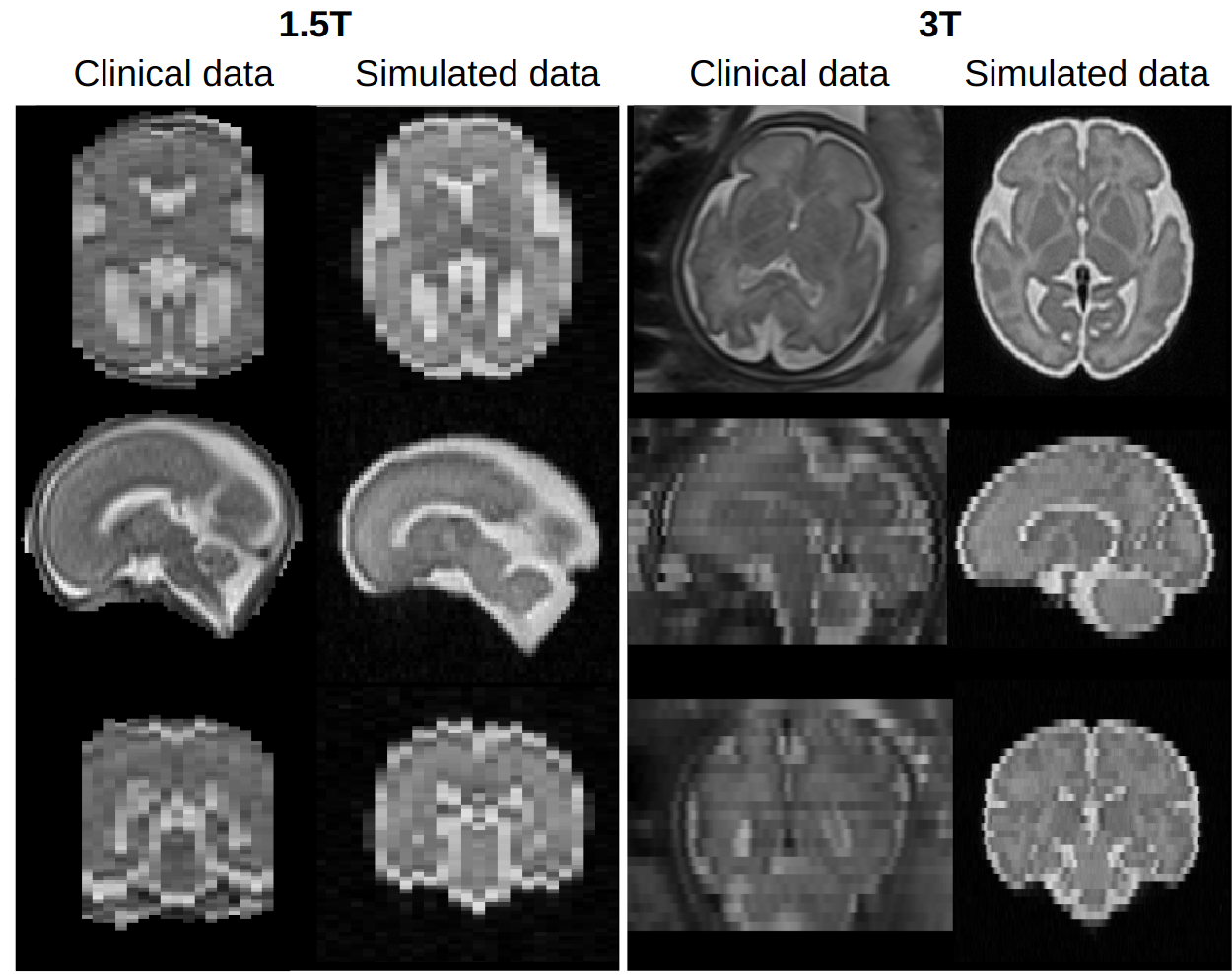}
  \caption{Comparison of simulated and clinical LR series at 1.5T ($1.1\times 1.1\times 3$) and 3T ($0.5\times 0.5\times 3 mm^3$) and 3T. }
  \label{fig:fabian_clinical}
\end{figure}

\end{document}